\documentclass[12pt,preprint]{aastex}

\input{epsf}

\begin{document}

\def\opeqn{\begin{equation}}
\def\cleqn{\end{equation}}

\def\lastrev{ Draft SHR\today}

\let\oldfootsep=\footnotesep
\setlength{\footnotesep}{.5\oldfootsep}

\def\lsim{\hbox{ \rlap{\raise 0.425ex\hbox{$<$}}\lower 0.65ex\hbox{$\sim$} }}
\def\gsim{\hbox{ \rlap{\raise 0.425ex\hbox{$>$}}\lower 0.65ex\hbox{$\sim$} }}

\def\etal{et~al.}
\def\kmskpc{\,{\rm km \, s^{-1} \, kpc^{-1}}}
\def\msun { \rm {M_\odot}} 
\def\angs {{ \rm \AA }}
\def\delchi{\Delta\chi^2}
\def\pac{Paczy{\'n}ski }
\def\ie{{\it i.e. }}
\def\thsky {\alpha}  
\def\vhat{\widehat{v}} 
\def\that {\widehat{t}}
\def\kms {\,{\rm km \, s^{-1} }}
\def\kpc {\, {\rm kpc}}
\def\kmsk {\,{\rm km \, s^{-1} \kpc^{-1}}}
\def\pri{^{\, \prime }}
\def\dpri{^{\, \prime \prime }}
%
\def\spose#1{\hbox to 0pt{#1\hss}}
\def\simlt{\mathrel{\spose{\lower 3pt\hbox{$\mathchar"218$}}
     \raise 2.0pt\hbox{$\mathchar"13C$}}}
\def\simgt{\mathrel{\spose{\lower 3pt\hbox{$\mathchar"218$}}
     \raise 2.0pt\hbox{$\mathchar"13E$}}}


\title { Can A Gravitational Quadruple Lens Produce 17 images? }
\author{ S.H.~Rhie   
}

\affil{Department of Physics, University of Notre Dame, Notre Dame, IN 46556 }

\vspace{5mm}
\begin{abstract} 
\rightskip = 0.0in plus 1em

Gravitational lensing can be by a faint star, a trillion stars of a galaxy,
or a cluster of galaxies, and this poses a familiar struggle between  
particle method and mean field method. In a bottom-up approach, a puzzle has
been laid on whether a quadruple lens can produce 17 images. The number of 
images is governed by the gravitational lens equation, and the equation for 
$n$-tuple lenses suggests that the maximum number of images of a point source 
potentially increases as $n^2+1$. Indeed, the classes of $n=1, 2, 3$ lenses 
produce up to $n^2+1 = 2, 5, 10$ images. We discuss the $n$-point lens system 
as a two-dimensional harmonic flow of an inviscid fluid, count the caustics 
topologically, recognize the significance of the limit points and discuss the 
notion of image domains. We conjecture that the total number of positive 
images is bounded by the number of finite limit points $2(n-1): n>1$ \ 
(1 limit point at $\infty$ if $n=1$). A corollary is that the total number 
of images of a point source produced by an $n$-tuple lens can not exceed 
$5(n-1):n>1$. We construct quadruple lenses with distinct finite limit 
points that can produce up to 15 images and argue why there can not be
more than 15 images. We show that the maximum number of images is bounded
from below by $3(n+1): n \ge 3$.  We also comment on ``thick Einstein rings" 
that can have one or more holes. 

\end{abstract}
\vspace{5mm}
\keywords{gravitational lensing }

\newpage
\section{Introduction}
\label{sec-intro}

Gravitational lensing phenomenon is ubiquitous throughout the universe and is 
poised to bear a substantial weight of the explorations of diverse objects 
from planetary systems \citep{gest} to the universe itself 
\citep{kai98,tys98,sch00}.
The utility of gravitational lensing dates back to the inception of Einstein's
general relativity (Einstein 1911, 1916; see Weinberg 1972; SEF 1992) 
whose first confirmation as the theory of gravity was offered by the measurements 
of the astrometric effect of the gravitational lensing by the sun of the 
background stars (Eddington 1919; see Weinberg 1972).    
The more dramatic lensing effect is the multiplicity of the images which 
arises due to the fact that there can be multiple null geodesics (photon paths)
that connect a distance radiation emission source and an observer because of the
focusing effect of the intervening lensing masses.

The first gravitationally lensed multiple image object to be discovered
was double quasar Q0957+561A,B \citep{wal79} whose two images 
separated by about 6 arcseconds plank the elliptical lensing (cD) galaxy 
located at about 5 arcseconds from the brighter image A.
A topological argument showed that a smooth extended mass distribution 
generates an odd number of images \citep{burke}\footnote{This paper has a 
special feature: It has only one page! In the case of $n$-point lenses,
Poincar\'e-Hopf index of Burke's vector field is $(1-n)$, and so the 
total number of images is even when $n$ is odd and vice versa.
Also, see section 2, and note the statement on logarithmic singularities 
in SEF, pp. 175.}, and so 
with the first double quasar surfaced the problem of missing images which 
may be too dim to be identified in the middle of the stars of the lensing 
galaxy or which may have been altered due to the hard nucleus effect of the 
stars of the lensing galaxy in which the assumption of continuum mass 
distribution has to be modified.

The effective lensing range of a point mass is given by the Einstein ring 
radius which is roughly the geometric mean of the Schwarzschild radius and 
the (reduced) distance of the lensing mass.  So, the lensing by a point mass 
is short-ranged and its singular nature can dominate the lensing behavior 
within its lensing range. The Schwarschild radius of a star is 
$\sim 10^{-5}$ (light) sec, the size of the
universe is $\sim 4\times 10^{17}$ sec, and so the Einstein ring radius of
a star at a cosmological distance is $\sim 2\times 10^6$ sec.  Since the 
typical distance between two stars in a galaxy is $\sim 10^8$ sec,
the effective number of stars that affect the local lensing behavior can
be a few to many depending on the local surface mass density of the galaxy
where the image under consideration is; 
the particle effect of the stars in a galaxy as a continuuum is ususally
discussed under the subject title of quasar microlensing because the 
stellar granularity 
\footnote{The effect of galactic granularity in a cluster lensing is of 
mesolensing where the astrometric shifts due to the galaxy nuggets can not 
be ignored. CL0024+1654 is a well-knwon example. As is common to any 
physical system, this intermediate scale phenomenon requires careful 
studies\citep{bro00}.}
mainly affects the photometric behavior rather than the 
global astrometric behavior or the  time delay \citep{you81}.  
Careful studies of the optical light curves of Q0957+561A,B 
have shown the absence of microlensing of time scales 40 - 120 days 
\citep{gil00}. The radio time delays of the double quasar at $4$cm and
$6$cm are somewhat inconsistent with the optical time delay $\approx 420$ 
days \citep{haa99}, and the residual systematics may be an indicator of 
radio activities to be learned. 

Quasar microlensing may be considered gravitational scintillation by a random 
distribution of an unknown effective number of point lenses \citep{you81}.
In a microlensing of a star, the lensing range of a microlensing star 
is order of $\sim 1$AU (the reduced distance of the lens is $\sim 10^6$ 
times smaller than in quasar microlensing), and so the gravitational
lensing involves only a single scattering and the lens is a low 
multiplicity $n$-point lens whose multiplicity is determined by the 
number of gravitationally bound objects within the lensing range of
the total lensing mass. These low multiplicity point lenses are particularly
interesting because the simplicity allows discoveries and measurements of the 
lensing systems which may be compact stars, planetary systems, or free-floating 
planets that are difficult to probe with other means. Observations of 
protoplanetary disks \citep{disk}, density wake fields, and dust rings  
\citep{dustring} are being actively pursued, and microlensing will offer 
extensive statistics on planets as the final products of the planet 
formations and evolutions. Microlensing will also offer extensive statistics
on stellar remnant black holes in the Galaxy adding our knowledge on the 
end point of the stellar evolution theories, the IMF, and the early 
history of the Galaxy.     

Microlensing is a relatively recent experimental phenomenon \citep{alc00} 
advanced with large format CCDs, fast affordable computers, high density data 
storage devices, and our total ignorance on dark matter species \citep{pac86}. 
The experimental necessity to interpret each lensing event exactly, as opposed 
to statistical nature of quasar microlensing \citep{qmicro}, affords  examining of
the low multiplicity point lens systems to exhausting details. We describe the 
simple behavior of the simplest lenses that is shared by arbitrary $n$-point 
lens systems:
the $n$-point lens equation is almost analytic and that imposes 
topological constraints on the caustic curves in equations (\ref{eqTop}) and 
(\ref{eqSum}). In the case of binary lenses, the contraints are strong
enough to specify all the topological configurations of the caustic curves
\citep{bin01}. With a bit of systematic toiling, one can also comb through 
the behavior of the class of triple lenses which is particularly useful for
multiple planetary systems \citep{tri01} or circumbinary planets 
\citep{97blg41}. 

In the following sections, we summarize the properties of the $n$-point lens 
equations, construct quadruple lenses that produce up to 15 images and argue 
that 15 is the maximum possible number of images of any quadruple lenses. 
We discuss the notion of image domains and illustrate the correlation between 
the limit points as the ``markers" of the positive image domains and the 
number of positive images of a point source. 
We conclude without a rigorous mathematical proof 
that the number of positive images of an $n$-point lens is bounded by the 
number of finite limit points $2(n-1): n>1$ and so the total number of images 
of a point source by an $n$-point lens can not exceeed $5(n-1): n > 1$. We 
leave it as a conjecture, having no shortage of margin, for an algebraic 
topologist or a combinatorian (perhaps with a finite Erd\"os number) who may 
find it interesting to complete the counting. We construct ``necklace" point 
mass configurations that have positive images at $(n+1)$ distinct limit points
to show that the lower bound of the maximum number of images is $3(n+1)$.
We comment on (thick) Einstein ring images in relation to the image domains 
of the interior of caustic curves.

\section { The $n$-point Lens Equation }

A gravitational lens of total mass $M$ consisting of $n$ gravitationally 
bound point masses whose interdistances are negligibly small compared to 
the distances of the center of mass to the positions of the observer ($D_1$) 
and the radiation emission source ($D_2$) 
is described by the following (two-dimensional linear) lens equation 
where $\omega$ is the position of the unlensed radiation 
emission source and $z$ is the position of an image of the radiation 
emission source generated by the point lens masses $\epsilon_j M$ 
located at $x_j: ~j=1, ... , n$.
\opeqn
 \omega = z - \sum_{j=1}^n {\epsilon_j\over \bar z - \bar x_j} \ ;
        \qquad
        \sum_{j=1}^n \epsilon_j = 1 \ .
\label{eqLens}
\cleqn
Here the lens plane where the two-dimensional position variables reside
and perpendicular to the line of sight of the (unlensed) source star 
passes through the center of mass of the lens elements; this choice is useful
when the focus is on the physics of the lensing system such as a microlensing
planetary system. The lens equation has been normalized to be dimensionless 
such that the unit distance scale of the lens plane parameterized by $\omega$ 
or $z$ is set by the Einstein ring radius of the total mass $M$ located at the 
center of mass.   
\opeqn
  1 = R_E = \sqrt{4GMD} \ ; \qquad  
   {1\over D} = {1\over D_1} + {1\over D_2} 
\cleqn
where $D$ is the reduced distance. (As is familiar from the reduced mass in 
mechanics, the reduced distance is smaller than the smaller of the two
distances $D_1$ and $D_2$.)  

The $n$-point lens equation (\ref{eqLens}) can be embedded in an ($n^2+1$)-th 
order analytic polynomial equation \citep{witt}, and so the number of images 
of a point source can not exceed ($n^2+1$) because an ($n^2+1$)-th order
analytic polynomial (a complex function of only $z$ or $\bar z$) always
has ($n^2+1$) solutions. A single lens produces two 
images, a binary lens produces three or five images,
a triple lens produces four, six, eight, or ten images, and  
it has been a puzzle whether a quadruple lens can produce 
up to $n^2+1 = 17$ images \citep{mao97}.  
The mininum number of images of an $n$-point lens is $n+1$ and one can count  
them easily by looking at the lens equation for $\omega = \infty$: there is
one positive image at $\infty$ and one negative image at each of the
$n$ lens positions. The smooth continuum mass in Burke's odd number theorem has 
one (unlensed) positive image near the source for large $\omega$.

\subsection {Linear Differential Properties of the Lens Equation}

The linear differential behavior of the lens equation 
plays an important role in lensing and is described by the Jacobian matrix 
of the lens equation.   	
\opeqn
   {\cal J} = \pmatrix{1 \  \bar\kappa \cr
                       \kappa \  1 } \ ; \qquad
           \kappa \equiv {\partial\bar\omega} 
               \equiv {\partial\bar\omega\over\partial z}
                    = \sum_{j=1}^n {\epsilon_j\over (z - x_j)^2}
\label{eqJacobian}
\cleqn
If an image of an infinitesimally small source is formed at $z$, the parity of
the image is given by the sign of the determinant $J$ of the Jacobian matrix
and the magnification of the image is given by $A = |J|^{-1}$.
\opeqn
   J(z) = 1 - |\kappa (z)|^2 \ ; \qquad J(z) \le 1
\cleqn
The set of solutions to $J(z)=0$ occupies a central position in lensing because
that is where the magnification $A(z)$ becomes large and the lensing signals
are most obvious. The lens equation is stationary where its Jacobian
determinant $J$ vanishes, and so the curve $J=0$ is called the critical curve.
Images flip the parity across the critical curve because $J$ changes its sign.
In fact, the images with opposite parities form pairs that are mapped to the
same source positions under the lens equation. 

Let's consider a real function $y = x^2$ for an easy illustration of the 
criticality: it is stationary or the linear derivative vanishes at $x=0$;  
two points   $x = \pm \delta x$ are mapped to one point $y = \delta x^2$; 
the Jacobian determinant $J_1$ of the $1\times 1$ Jacobian matrix ${\cal J}_1$ 
is $J_1 = {\cal J}_1 = dy/dx$,   
and $J_1 >0$ for $x = \delta x $ ($>0$) and  $J_1 <0$ for $x = - \delta x$;
$y =0$ on which the critical point $x =0$ is mapped under the real function
divides the $y$-axis into two regions: $\{y| y > 0\}$ where each point $y$ has 
two solutions $x = \pm\sqrt{y}$ and $\{y| y > 0\}$ where each point $y$ has no
solutions. The set of points onto which the critical curve is mapped under
the lens equation is called the caustic curve and each caustic point defines 
a direction (transverse to the caustic curve) that is locally analogous to 
the $y$-axis where the caustic point is at $y=0$.

In the neighborhood of a caustic point, 
point sources on one side of the caustic point (say $y>0$)
produce pairs of images with opposite parities across the corresponding 
critical point (analogous to $x=0$) while point sources on the other side 
of the caustic point (say $y<0$) do not. Thus, the caustic curve defines the
boundary across which the number of images changes by two (or a multiple of
two at the intersection points). In other words, in the neighborhood of the 
critical curve the lens equation is two-to-one, and {\it two images appear from
or disappear into the critical curve as the source crosses the caustic curve.}
The italicized statement is well-known, and the focus of this paper is to 
discuss what is underlying the behavior of caustic curves as the boundaries 
of the caustic domains  and  to convey that caustic curves are 
easily comprehensible benign curves despite the  
hostile appearance of the spikiness near the cusps. It all starts from a simple 
observation that the Jacobian matrix is determined by one analytic function 
$\kappa(z)$.  The lens equation is a function of both variables $z$ and $\bar z$, 
but the derivatives are (anti)analytic. So, complex coordinates are useful
for the $n$-point lens equation as are spherical coordinates for spherically 
symmetric systems.

\begin{figure}
\plotone{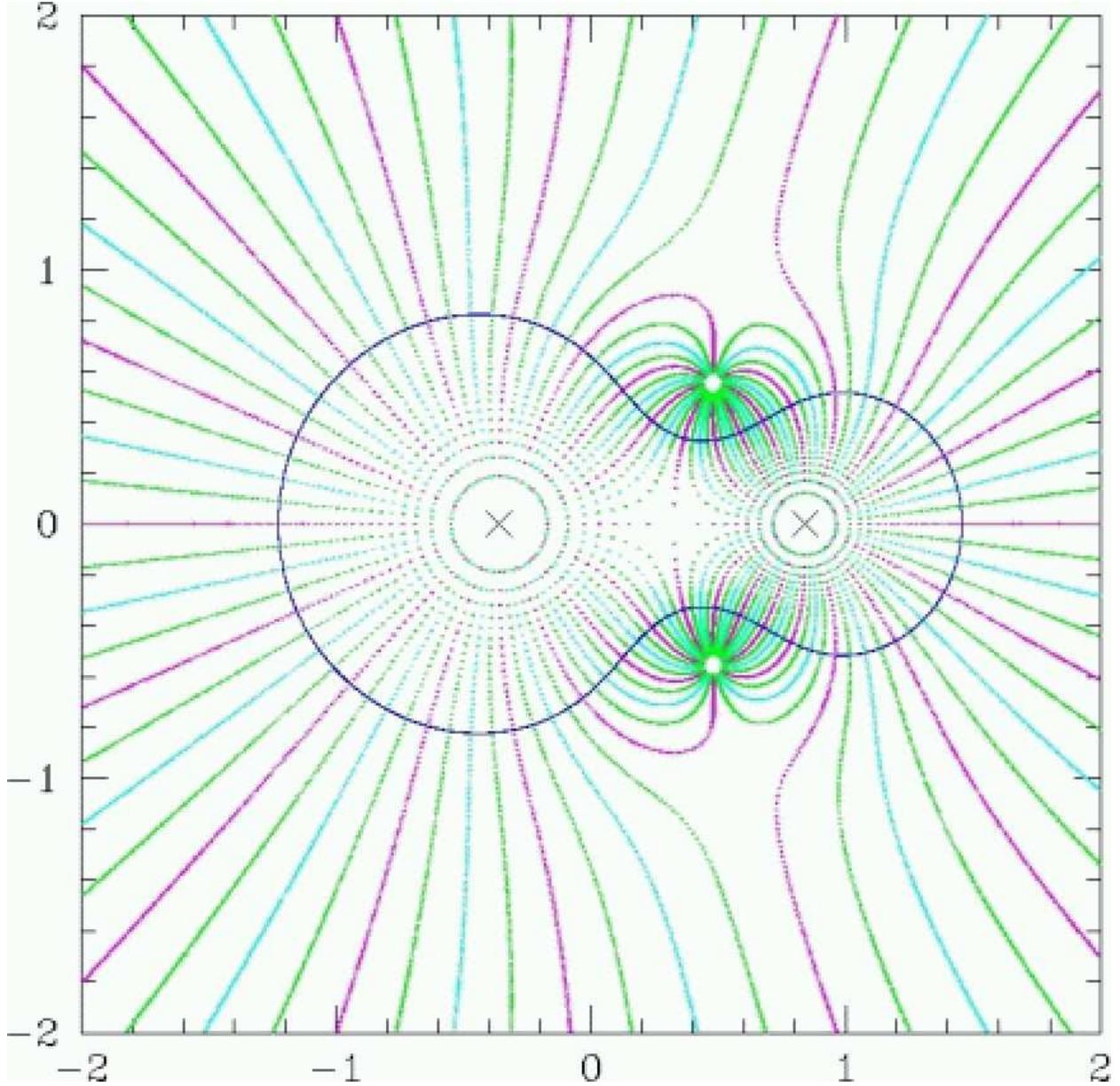}
\caption{The field lines emanating from the lens positions marked by X flow
  into the limit points and infinity. The mass fraction of the smaller
  mass on the right is $\epsilon_2 = 0.3$, and the same number of field
  lines emerge from each lens position. The field lines from the dominant
  mass sweep the larger part of the lens plane. The closed curve is the 
  critical curve and is orthogonal to the field lines. Three bifurcation
  points can be seen: one on the lens axis and two off the axis.   
  \label{fig-kf} }
\end{figure}

\underline{\it The Analytic Function $\kappa(z)$ Defines a Potential Flow 
on the Lens Plane.} \ \
An analytic function describes a two-dimensional harmonic flow of an inviscid
fluid whose streamlines are conserved except at the ``sources" and ``sinks" 
given by the poles and zeroes of the analytic function \citep{landau}.
\opeqn
  \bar\partial \kappa = 0  \qquad \Longleftrightarrow \qquad
  \nabla^2 \kappa = 0 
\cleqn
The $n$ lens positions are the poles of $\kappa$ and are double poles.
In the neighborhood of a lens position,  $z \approx x_j$,
\opeqn
  \kappa \approx {\epsilon_j\over (z - x_j)^2} \ .
\label{eqPole}
\cleqn
The zeros of $\kappa$ are referred to as (finite) limit points because 
$J$ takes the maximum value $1$ where $\kappa = 0$. 
Equation (\ref{eqJacobian}) shows that there are $2(n-1)$ limit points
and they are simple zeros unless degenerate.  If $z_\ast$ is a limit point, 
near the limit point $z \approx z_\ast$,
\opeqn
  \kappa  \approx \partial\kappa(z_\ast) (z - z_\ast) \ . 
\label{eqZero}
\cleqn
The infinity behaves as a double zero. If $z \rightarrow \infty$, 
\opeqn
  \kappa \approx \sum_1^n {\epsilon_j\over z^2} = {1\over z^2} \ .
\cleqn  
If the phase angle of $\kappa$ is $2\varphi$,
\opeqn
  \kappa = |\kappa| e^{2i\varphi} \ , 
\label{eqKappa}
\cleqn
then $\varphi$ changes by $\pi$ around each (finite) limit point, 
$2\pi$ around the infinity, and $-2\pi$ around each lens position where the 
positive orientation is counterclockwise as usual. 
So, the phase angle of $\kappa$ defines the stream lines or field lines that 
flow from the lens positions to the finite limit points and infinity.
See figure \ref{fig-kf} for the field lines of a binary lens. 
The magnitude $|\kappa|$ behaves as the potential
and decreases monotonically along the field lines from $|\kappa| = \infty$
at the lens positions to $|\kappa| = 0$ at the limit points. 
The equipotential curves ($|\kappa|=$ constant) are smooth because the lens
equation is smooth everywhere except at the poles ($\kappa = \infty$) and
are orthogonal to the field lines ($\varphi =$ constant). The total 
number of field lines that cross an arbitrary equipotential curve 
($|\kappa|=$ constant) which may consist of multiple loops is 
$\Delta \varphi = 2\pi n$.   

\underline{\it 
The Critical Curve is but an Equipotential Curve given by $|\kappa| = 1$.} \ \  
The critical curve is an equipotential curve (with the potential value $1$)
since $|\kappa| = 1$ where $J=0$, and it is easy to visualize the configurations
because of the orderly behavior of the family of equipotential curves.
If we consider the lens plane as a large
two-sphere (with one point $\infty$ at infinity -- the well-known notion 
of one-point compactification), the equipotential curves and field lines
can be considered latitudes and longitudes of a multiple-pole sphere with 
the north poles at the lens positions (``sources") 
and the south poles at the limit points and infinity (``sinks"). 
Check the critical curves in this paper for the 
predictable configurations. The normals to the equipotential curves and
field lines form a right-handed basis vector field.
\opeqn
  \{\nabla|\kappa|, \nabla\varphi\} 
  \Leftrightarrow 2 \{\bar\partial|\kappa|, \bar\partial\varphi\} 
\cleqn 
The multiplicity of the ``sources" and ``sinks" implies that there are ``forks"
in the streamlines where the equipotential curves bifurcate. The bifurcation
points (or saddle points: ~$\partial\kappa =0$) are four-prong vertices 
(or $\times$-points), and the basis vectors are not uniquely defined at the 
bifurcation points. There are three bifurcation points in every binary lens 
\citep{bin01}. For higher $n$ ($ >2$), the number varies with the lens parameters.      

A critical loop (a connected closed curve with $J=0$) can be assigned an 
integer or a half-integer because of the analyticity of $\kappa$. 
If the closed curve encloses $N_\times$ lens  positions and $N_\ast$ limits
points, we define the associated topological charge $e$ as follows.
(It is valid for an arbitrary closed curve that is analytically connected 
to the critical curve: the residue theorem.)
\opeqn
  e = N_\times - {N_\ast\over 2}
\cleqn
Then the net flux of the field lines that flow out through the critical loop 
is $\Delta\varphi = 2\pi e$. The charge of a limit point is negative 
($e = -1/2$) because the field lines flow in while the charge of a lens
position is positive ($e = 1$). The relative sign reflects the opposite
orientations of the critical loops around a limit point and a lens position,
which derives from the analytic behavior of $\kappa$ around a zero 
and a pole in equations (\ref{eqZero}) and (\ref{eqPole}).  The critical 
curve of a lens can consist of many loops, and the topological charges 
associated with the critical loops are subject to a constraint because
the total number of field lines is determined by the number of lens 
elements $n$: ~$\Delta\varphi = 2\pi n$.
\opeqn
  \sum_j |e_j| = n
\label{eqSum}
\cleqn

\underline{\it 
If $z$ is a Critical Point, $\omega(z)$ is called a Caustic Point.} \ \ 
Since the lens equation is smooth (and so continuous) in the neighborhood of 
the critical curve, the caustic curve ~$\{w(z)| \ |\kappa(z)| = 1\}$ shares
the connectivity of the critical curve and is smooth except at the 
stationary points where cusps form.  The caustic curve is made of the same 
number of loops the critical curve is made of and bifurcates where the 
critical curve bifurcates.  So a caustic loop is 
associated with the same topological charge of the corresponding critical
loop, and we will see that the charge $|e|$ measures the smooth rotation of 
the tangent of the caustic loop. At the cusps, the tangents change the 
directions abruptly by $\pi$. The critical points that are mapped to cusps
under the lens equation are called {\it precusps}. 

In the case of a single lens (Schwarzschild lens), every point on the critical 
curve (Einstein ring) is a precusp, and the point caustic may be considered 
a degenerate cusp. If we look at the caustics in this paper, the cusps are 
just a finite number of punctuations on the otherwise smooth curves. So our
guiding thought for an intuitive or impressionistic understanding (or
handwaving) would be: How badly are the critical curves of $n$-point 
lenses deviated from the circular Einstein ring of a single lens?   
If we look at the critical curves in this paper, we do find some
comforting similarities in the segments around lens positions. That is where
the critical curve is more or less tangential to the critical direction 
($E_-$ defined below).	The segments around limit points include parts
where the critical curve is more or less normal to the critical direction.   
If we look at the critical curve in figure \ref{fig_idomain}, its 
(extrinsic) curvature sign changes around the precusps corresponding to
the off-the-axis cusps, but $E_- = i e^{-i\varphi}$ smoothly rotates 
with increasing $\varphi$ along the critical curve. So, it should not be 
surprising that the connected critical curve behaves differently around the 
limit points from around the lens positions.

The critical condition $J=0$ implies that (at least) one of the eigenvalues 
of the Jacobian matrix ${\cal J}$ vanishes on the critical curve. 
The $2\times 2$ matrix ${\cal J}$ has two eigenvalues
\footnote{The Jacobian determinant is the product of the two eigenvalues, 
and $J=0$ result in two possible types of critical curves: 
``tangential critical curve" where the critical curve is
tangent to the critical direction ($E_-$; see section 2)
and ``radial critical curve" where
the critical curve is normal to $E_-$ (SEF pp231). A single
(Schwarzschild) lens has only a ``tangential critical curve" which is
the Einstein ring (the second factor of $J$ is positive and $\ne 0$)
and each point of the ``tangential critical curve" is a precusp.
In general, the critical curve of an $n$-point lens is a mixture of the
parts that are more or less tangential to $E_-$ and parts that are more
or less normal to $E_-$.}, 
and $\lambda_-$ vanishes on the critical curve.
\opeqn
  \lambda_\pm = 1 \pm |\kappa|   
\cleqn
The eigenvectors ($\pm e_\pm$: blind to the senses) are conveniently described 
by the basis vectors ($e_\pm$) (with definite orientations).
\opeqn
 e_+ \equiv {e^{-i\varphi} \choose e^{i\varphi}} \ , \quad
 e_- \equiv {i~e^{-i\varphi} \choose -i~e^{i\varphi}}  \ ;
 \qquad ||e_\pm|| = \sqrt{2} 
\cleqn
An arbitrary vector $dz$ can be decomposed in terms of the 
basis vectors: $dz = dz_+ E_+ + dz_- E_-$ where $dz_\pm$ are real
and $E_\pm$ denote the upper components of $e_\pm$.

\underline{\it
The Critical Direction is $\pm E_-$.
The Caustic Curve is Tangent to $\pm E_+$.} \ \ 
Let $dz$ be an arbitrary displacement from a critical point $z_c$,
then $\lambda_-(z_c) = 0$, and $d\omega = dz_+ \lambda_+ E_+ = 2 dz_+ E_+$.
So the lens equation is critical or stationary in the eigendirection 
$\pm E_-$: ~if $dz_+ =0$, then $d\omega = 0$.  If $dz = dz_- E_-$, the 
lens equation restricted to the critical direction is quadratic \citep{limb}.
\opeqn 
  \delta\omega_{2c}  = {1\over 2}\bar\partial J ~dz_-^2
 = {1\over 4} ~dz_-^2 ~\left(\partial_+ J E_+ + \partial_- J E_- \right) \ ,
\label{eqQuadc}
\cleqn
where $dJ =  dz_+ \partial_+ J + dz_- \partial_- J$. In an exact analogy to
the real function $y = x^2$ we discussed above, the equation (\ref{eqQuadc})
has two solutions ($\pm dz_-$) only if $\delta\omega_{2c}/\bar\partial J >0$.
Then, the remaining question is: Where is $\bar\partial J$ pointing? In the 
$z$-plane, it is a normal vector to the critical curve (or any equipotential
curve): it is normal outward to the critical curve with positive topological
charge $e>0$, and normal inward to the one with $e < 0$. In order to determine
the relative direction of $\bar\partial J$ in the neighborhood of the caustic
curve (in the $\omega$-plane), we need to find a reference direction that is
well defined intrinsically. We first note that the caustic curve is always 
tangent to the eigendirection $\pm E_+$. So the normal component
$(\partial_- J E_-)$ determines on which side of the caustic curve a point
source at $\omega(z_c)+ \delta\omega_{2c}$ produces two images at $z_c\pm dz_-$   
and we can use the (extrinsic) curvature vector of the caustic curve as the 
reference direction.   

\begin{figure}
\plotone{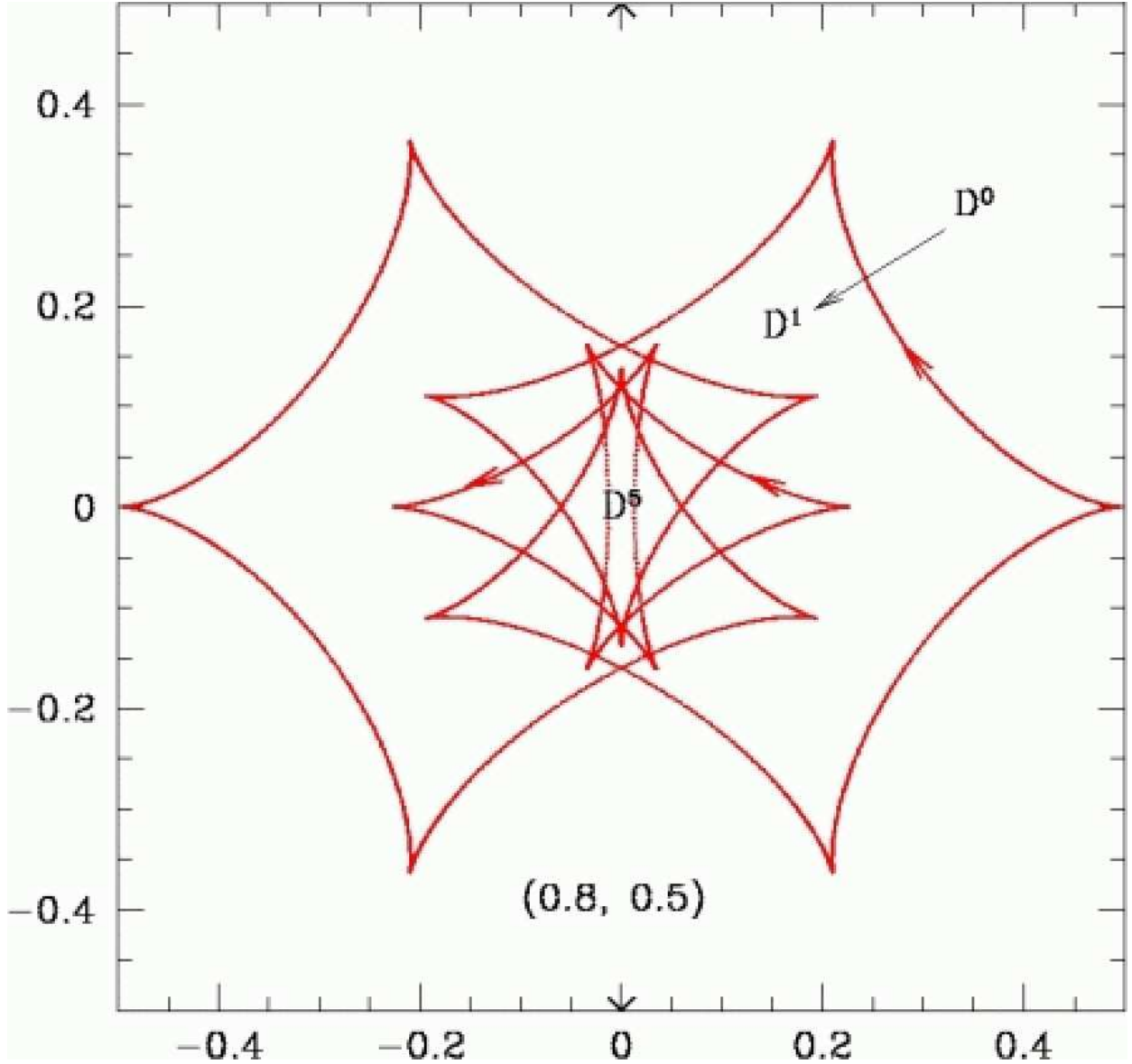}
\caption{Four point masses are at $x_{1,3} = \pm 0.8$ and 
   $x_{2,4} = \pm 0.5i$. The caustic curve consists of 
   three caustic loops: $(q,m) = (3,12), 
  (1,3), (1,3)$. The arrows show the orientation of the caustic loops
  ($d\varphi > 0$). $D^0$ is the lowest degree domain where the number of
  images is 5, and the domain $D^1$ generates 7 images. 
  $D^0 \longrightarrow D^1$ depicts the direction in which the number of
  images increases by two in relation to the orientation of the caustic loop.  
  All the caustic loops wind around in the same direction -- counterclockwise. 
  Crossing through an intersection point changes the degree of caustic
  domain by two and the number of images by four. In this lens configuration, 
  there are no intersection ponts where more two curves intersect.
  A point source in $D^5$ generates 15 images. 
  \label{sym4-caus3} }
\end{figure}

If $dz = dt$ is tangent to the critical curve ($d\varphi >0$), $d\omega (z)$
is tangent to the causic curve and is in the direction of 
$(\partial_- J E_+)$. 
\opeqn
\delta\omega_1 = 2 dt_+ E_+ 
   = {|dt|\over |\bar\partial J|} ~\partial_- J E_+    
\label{eqCaus}
\cleqn
The second order term shows that the caustic curve bends in the direction of 
$ (-\partial_- J E_-)$. 
\opeqn
\delta\omega_{2t}  = {1\over 2} \bar\partial \bar\kappa ~dt^2
  =  {1\over 4} |dt|^2 ~(\partial_+ J E_+ - \partial_- J E_-) 
\label{eqQuadt}
\cleqn
So, if $\pm E_+$ define the $x$-axis such that the caustic curve is 
convex upward, then the quadratic equation (\ref{eqQuadc}) has two solutions 
where $y \equiv \delta\omega_{2c} > 0$ and none where $y < 0$.     
We also observe from (\ref{eqCaus}) and (\ref{eqQuadt}) that the tangent to 
the caustic curve rotates clockwise ($d\varphi > 0$) unless $\partial_- J =0$. 
See figure \ref{sym4-caus3} for an illustration. So, in order to form a caustic 
domain that produces many images, the caustic curve will have to wind around
many times (here always counterclockiwise) intersecting itself and nesting.   
The caustic curve in figure  \ref{sym4-caus3} has total winding number 5.  

The lens equation restricted to the critical curve is stationary 
($\delta\omega_1 =0$) where $\partial_- J =0$, and the caustic curve develops 
a cusp.  If $m$ is the number of the cusps of a caustic loop, the tangent 
of the caustic loop rotates by $-2\pi |e|$ (clockwise) in total on the smooth 
segments and flips by $m\pi$ at the cusps resulting in the net rotation of 
$ 2\pi q = m\pi - 2 \pi |e|$ where $q$ is the winding number.    
\opeqn 
  q = {m\over 2} - |e|   
\label{eqTop}
\cleqn
A caustic loop with $q \ge 2$ intersects itself and can form a caustic domain   
where a point source generates $2q$ more images than outside the caustic loop. 
$|e|$ is easily determined from the corresponding critical loop and subject to 
the constraint $\sum_j|e_j| = n$ in equation (\ref{eqSum}). 
Caustic loops intersect and nest each other, 
and the number of images jumps by two at each crossing of the
caustic loop. The case $q=1$ is like an $m$-gon (polygon with $m$ vertices)
and $|e|$ amounts to the total internal angle in units of $2\pi$.   
A critical loop enclosing a limit point has $|e| = 1/2$, and so the corresponding 
caustic loop has ``internal angle" $\pi$ and is triangular:  $(q, m) = (1, 3)$.
The caustic curve of a quadruple lens in figure \ref{sym4-caus3} consists of 
two triangular caustics and one with 12 cusps: $(q, m) = (3, 12)$, and it 
satisfies the constraint $\sum_j|e_j| = 1/2 + 1/2 + 3 = 4$.    

A corollary will be the {\it theorem of even number of cusps} (SEF, pp213).  
From equations (\ref{eqTop}) and (\ref{eqSum}), 
\opeqn
  \sum_j{m_j\over 2} = n + \sum_j q_j \ . 
\cleqn
Since the RHS is an integer, $\sum_j m_j$ is an even number. 
Thus in an $n$-point lens, the number of caustic loops with odd number of 
cusps is always even. In a binary, trioids: $(q, m) = (1, 3)$ appear in a
pair that are reflection symmetric with respect to the lens axis. The triple
lens in figure 1 in \cite{infimum} consists of two caustic loops with odd
number of cusps: $(q, m) = (1, 3)$ and $(2, 9)$. The quadruple lens in 
figure \ref{sym4-caus3} includes two trioids.

\section {Can a Quadruple Lens Produce 17 Images?}

The fact that there are triple lenses that can produce up to 10 images
was discovered while we were investigating the location of a point
source that produces images at the limit points \citep{infimum}. 
There are four (finite) limit points in a triple lens, and the
number of images of the point source that produces images at the four limit
points must be at least ten because there are always two more negative images
than positive images. Since a triple lens can produce no more than 10 images,
a correlation between the maximum possible number of images and the limit points
was suspected. We found that a point source can produce an image at each of
$2(n-1)$ limit points only in the cases of $n= 2$ and $n=3$. 

\begin{figure}
\plotone{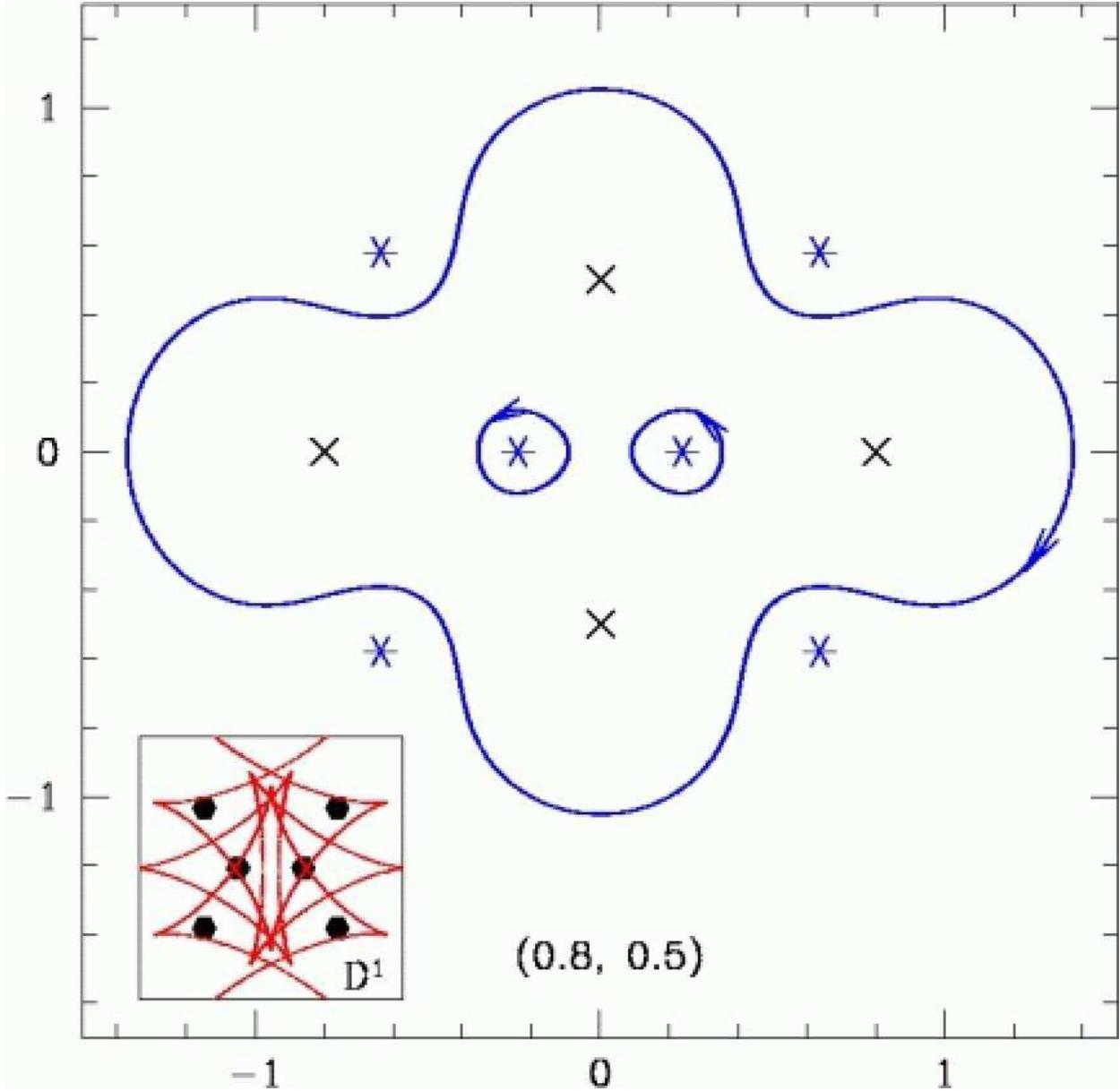}
\caption{The critical curve for the lens in figure 2 
  consists of one large loop with $e = 3$ and two small loops each with
  $e = -1/2$. The 6 finite limit points ($z_\ast$) are marked by $\ast$'s, 
  and their corresponding positions ($\omega_\ast$) in the $\omega$-plane 
  are shown in relation to the caustic curve in the inset. None of the 
  limit points are the image positions of a source in the domain $ D^5$. 
  \label{sym4-crit3} }
\end{figure}

If a quadruple lens were to produce 17 images with all the positive
images at the limit points, there would have to be 7 limit points because 
there are always three more negative images in a quadruple lens.
A quadruple lens has only 6 (finite) limit points. We find quadruple lenses with 
6 distinct (or non-degenerate) limit points that can have caustic domains 
${\cal D}^5$ where a source produces 15 images (Fig. 2, Fig. 7),  
and the limit points ($z_\ast$)
are not mapped onto one point in the $\omega$-plane (Fig. 3).
We can see from figure \ref{sym4-caus3} and figure \ref{sym4-crit3} that one 
extra limit point and associated $|e|=1/2$ caustic curve could make a domain 
${\cal D}^6$ with 17 images, but that would violate the constraint 
$\sum_j|e_j| = 4$ (and the theorem of even number of cusps). 

\begin{figure}
\plotone{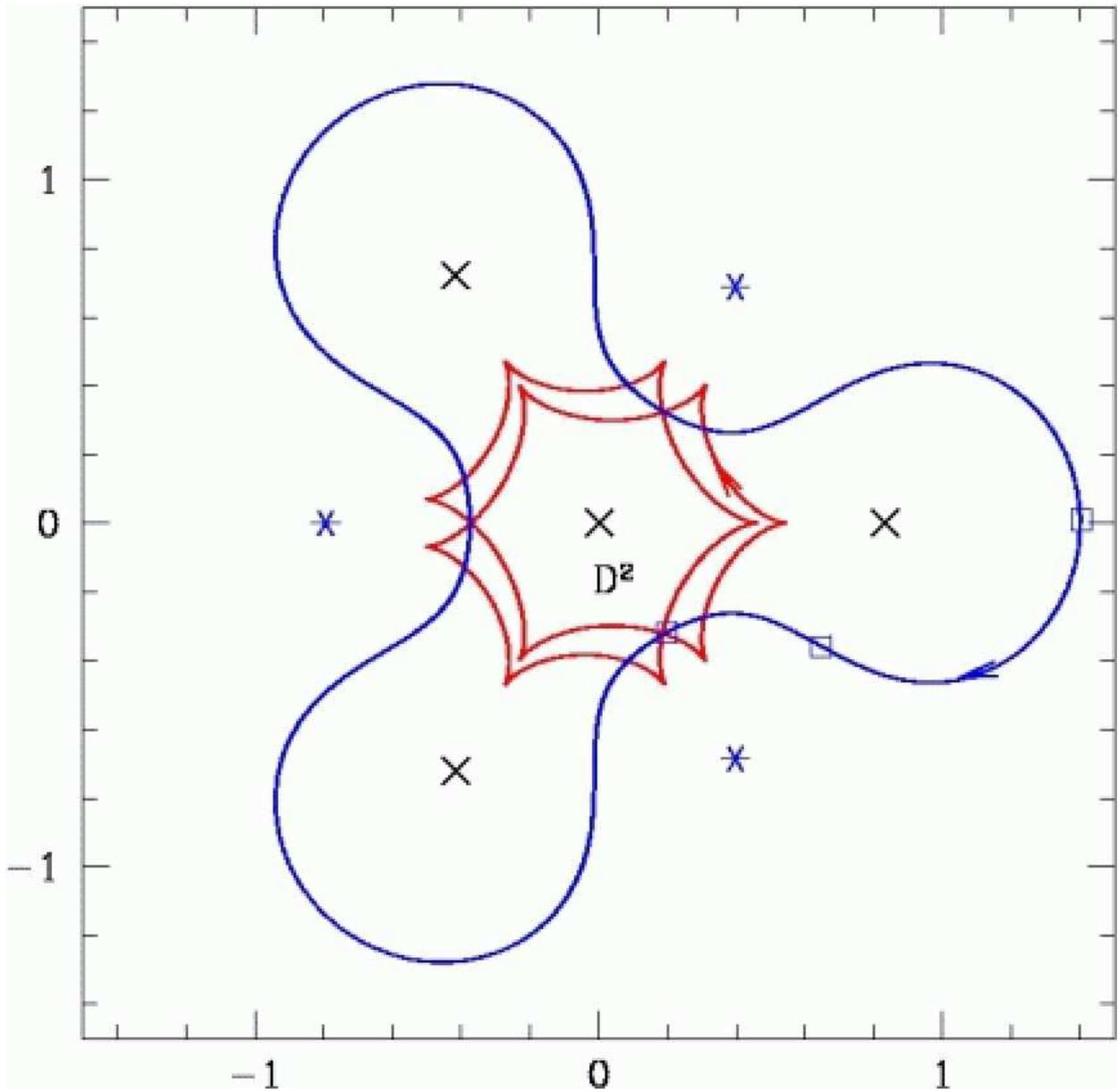}
\caption{The highly symmetric triple-like quadruple lens has only 3 distinct 
  limit points because each one is doubly degenerate. Three {\it precusps}  
  are marked by open squares, and the other nine can be easily identified 
  from the three through the symmetry. The three precusps near the center are 
  notable because they reflect the equal strengths of the topological charges 
  of the degenerate limit points ($e=-1$) and the lens positions ($e=1$).      
  The critical can be considered to consist of six segments each of 
  which is concave toward one of the three lens posistions off the center 
  or one of the three limit points and marked by three precusps.
  \label{fig-sym3} }
\end{figure}

The quadruple lens in  figure \ref{fig-sym3}
is triple-like and allows a source position that has an image at every limit
point. The limit points are (doubly) degenerate due to the high degree of symmetry
of the lens configuration, and the caustic curve is relatively simple accommodating
only up to ${\cal D}^2$ where a source generates 9 images (3 positive images
and 6 negative images); a source at the center produces 3 images on each 
three-fold symmetry axis.  If the three lens elements off the center are 
on a circle of radius $a$, the 3 positive images of the source at the center
are formed at the limit points when $a = 2^{-{1\over 6}} = 0.89$. 

\begin{figure}
\plotone{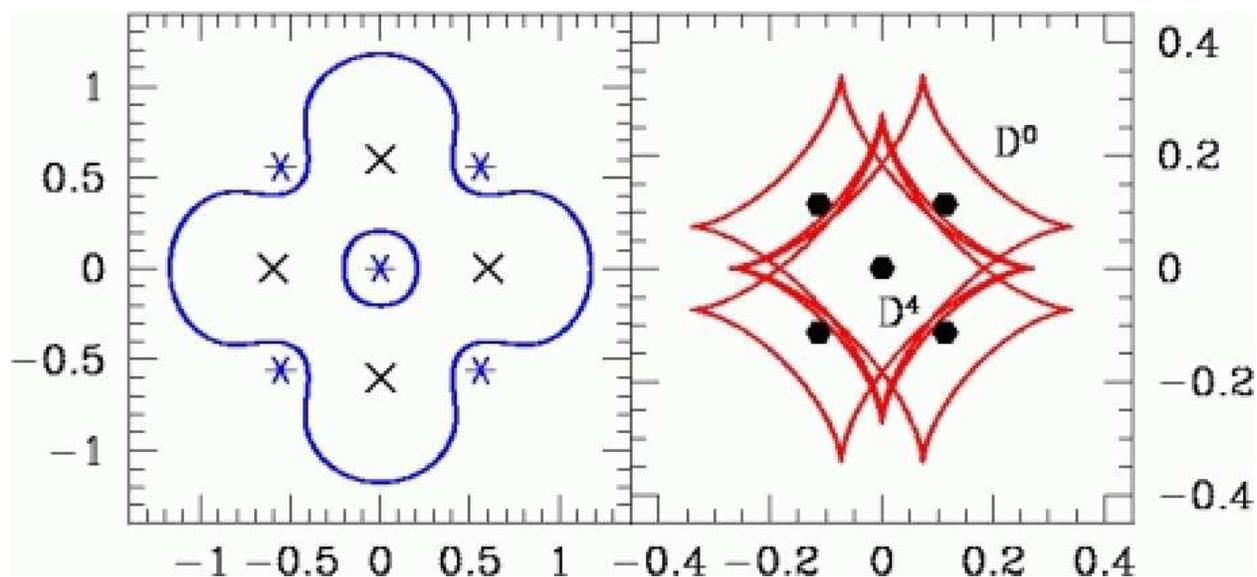}
\caption{Four equal mass lens elements are equally spaced on a circle 
  of radius $a = 0.6$. The four-fold symmetry forces the limit point at 
  the center to be degenerate. The small critical loop at the center has 
  $e = -1$, and the corresponding caustic loop with four cusps can be
  seen on the right panel. Each lens position claims a segment of the
  large critical loop each with three precusps. The caustic curve with
  $|e|=3$ has total 12 cusps, and four of them are hard to distinguish 
  from the four cusps of the caustic loop $|e|=1$. The highest degree
  caustic domain is ${\cal D}^4$ where a point source generates 13 images.        
  When $a = 3^{1\over 4}/2 = 0.658$, a point source at the center of mass
  generates 5 positive images at the limit points.    
  \label{fig-sym4} }
\end{figure}

The quadruple lens in figure \ref{fig-sym4} has five
distinct limit points because the one at the center is doubly degenerate, and
the highest degree caustic domain is ${\cal D}^4$ where a source generates
13 images. As the two limit points at the center in figure \ref{sym4-crit3}
merges, the two intersecting triangular caustics in figure \ref{sym4-caus3}
``merge and unwrinkle" into a quadratic caustic in figure \ref{fig-sym4}. 
An intermediate step after the ``merger" is shown for two cases in figure 
\ref{fig-sym4-ab1}: the domains ${\cal D}^5$ are split and diminished in size.    
If we consider a family of lenses in which four equal mass lens elements 
are equally spaced on a circle of radius $a$ (such that $x_1 = a$), 
the source at the center of mass generates 5 positive images at the 
limit points when $a = 3^{1\over 4}/2 = 0.658$. The total number of images 
is 13, and the source is in ${\cal D}^4$ which is the highest degree caustic
domain for the family of regular rhombus lens configurations.

What is implied is that the maximum number of images can be obatined when
the configuration of the lens elements is largely symmetric to maximize the
gravitational interference of all the lens elements as a whole but not so 
symmetric as to generate degenerate limit points.  
When the lens elements are dispersed, the gravitational interference is 
fractionized and the lens system effectively behaves like a linear sum of 
lower number of lens systems (Fig. \ref{fig-sym3lim}).
When the limit points are degenerate, each degeneracy costs 2 images.
In fact, the limit points do not have to be degenerate to cost images:
When they are close enough, the critical loops around two limit points 
merge to form one connected positive image domain. The residual highest
degree domains ${\cal D}^5$ in figure \ref{fig-sym4-ab1} disappear  
quickly when the lens configurations become more regular. 
One trivial nontheless worthy point to note is that a degenerate limit 
point can host only one image of a point source (property of the lens
equation) even though the degeneracy has its direct bearing on the 
$e$-charge of a loop (such as critical loop) enclosing it (linear 
differential behavior of the lens equation). Then, the significance of the 
limit points must lie in that they are the ``markers" of positive image 
domains, which we will conjecture to be the case after briefly examining 
the notion of image domains.

\underline{\it The Images of the Caustic Curve Defines the Image Domains.} \ \ 
We are convinced by now that the caustic curve of a lens defines the caustic 
domains, and it is clear what we must have meant by the fact that the caustic 
domain  ${\cal D}^5$ of the quadruple lens in figure \ref{sym4-caus3} generates 
15 images. There are 15 image domains in the image plane whose boundaries are 
mapped onto the boundary of the ${\cal D}^5$ under the lens equation. The image
plane with the domain boundaries of the quadruple lens is a bit of mess to use 
for an illustration even though we can say that it is rather pretty. 
So, let's examine the class of binary lenses 
that offers a simple laboratory case for the notion of 
image domains; the caustic curves are simple curves (winding number $q=1$)
and there are only two types of caustic domains: {\it outside} for 3 images
and {\it inside} for 5 images. The fact that the caustic curve is sandwiched
between the domains of 3 images and 5 images implies that the caustic curve itself
generates 4 images. In other words, there are 4 curves in the image plane that 
are mapped onto the caustic curve, one of which is obviously the critical curve. 
The critical curve is smooth as we are all familiar with by now, but the others
do have cuspy points. The lens equation at the cuspy points is smooth without
criticality, and so the corresponding cusps (kinky points) on the caustic curve 
generate kinky points in the $z$-plane.   

Figure \ref{fig-idomain} shows the image domains of a binary lens with
a connected caustic with six cusps that are defined by the 4 images
curves of the caustic curve shown in the inset. Smooth one is the critical curve,
and the other 3 curves are tangent to the critical curve at the precusps
(each curve is attached to the critical curve at two precusps). The 3 curves 
are smooth at the precusps on the critical curve where the lens equation 
(restricted to the critical curve) is stationary, hence each of the 3 curves 
have four cuspy points that are the images of the cusps of the caustic curve 
where the lens equation is non-critical. The four curves that are the image 
curves of the caustic curve form 5 image domains around the critical curve.
They are where the 5 images of an emission source inside the caustic curve
form and consist of 3 negative image domains (labelled ${\cal D}^1_-$) and
2 positive image domains (${\cal D}^1_+$). The limit points are marked by
$\ast$ as is the case throughout this paper, and they are inside the two
positive domains ${\cal D}^1_+$.  The entire lens plane is divided 
into 8 domains by the image curves of the caustic curve: 
$5\ {\cal D}^1$'s and $3\ {\cal D}^0$'s, and they account for 3 images for
a source outside the caustic curve and 5 images for one inside.    
If we consider our Local Group as a binary lens made of the Galaxy and M31
at a cosmological distance, an elliptical source galaxy can fill the
{\it inside} caustic domain to produce a large ring image with two holes
which we may refer to as ``thick Einstein ring" with two holes.

In the neighborhood of a precusp the image domains are divided into four 
sections, which is related to the fact that {\it the lens equation restricted 
to the critical direction is cubic at a precusp and produces 3 solutions when 
the source is inside the caustic and 1 solution when the source is outside}.
The four domains around a precusp consist either of 3 positive image domains 
and 1 negative image domain (say, type $+$), or of 1 positive image domain and 
3 negative image domains (type $-$). The two cusps on the lens axis are of 
type $+$, and the four cusps off the lens axis are of type $-$. When a source 
crosses the caustic through a cusp of type $+$~ ($-$), two positive (negative)
images are created and move in a tangential direction to the critical curve, 
and one positive (negative) image crosses the critical curve changing
its parity. The net parity remains the same as it does when the caustic
crossing is away from a cusp. It should be worth reemphasizing that the 
well-known statement that {\it two images of opposite parities appear or 
disappear at a caustic crossing} which is strictly valid for line caustic
crossings has to be relaxed for cusp crossings because the pair of images
that appear or disappear have the same parities (rh97). 

These microscopic
discussions which may seem irrelevant or annoyingly nitpicking fit together
rather naturally once we draw the image domains, and simple intuitive 
interpretations fall out: A source near a cusp inside the caustic has 3
highly magnified images because they are all near the critical curve; if 
the source moves along a line caustic, two images remain large because they
are near the critical curve and the 3rd one shrinks because it is away from
the critical curve; the 3rd image did not come into the picture as the other
two small images when we considered the quadratic equation for two large 
images in equation (\ref{eqQuadc}) because of its distance from the critical
curve. The mysterious behavior of the parities of the two images that appear
or disapper across the caustic curve we discussed above (and once was a 
source of caustic comments from referees) is just a {\it partner swapping}
process among the three highly magnified images whose process one can 
comprehend easily by considering a source moving along the caustic curve
past a cusp; the caustic curve near a cusp is reflection-symmetric with respect 
to its tangent vector (eigenvector $E_+$) at the cusp, and a source approaching 
the cusp along the symmetry axis from inside the caustic generates two images 
that are reflection-symmetric with respect to the symmetry axis and one image 
on the symmetry axis; the two images off the symmetry axis ($E_+$) are on
the same side of the critical curve and so have the parity.

Figure \ref{fig-idomain2} shows the image domains of a case where the 
caustic curve with six cusps is split into two quadroids. The long arrow 
indicates 3 large image curves of the 4-cupsed caustic centered around 
$(0, 1)$.  If a quasar with its host galaxy covers this caustic loop,
there will be four images of the quasar and the finite size galaxy will
form a ring image with one hole. Two of them are postive images and the
other two are negative images. The fourth image curve of the caustic is inside 
the the image ring of the other caustic loop and is marked by the shorter arrow.  
This small image domain generates the fifth image (negative image) and has 
a high probability to become a missing image because of its smallness. 
What we note is that when the critical curve splits, the number of positive 
image domains for each caustic loop is never more than the number of positive 
image domains of the connected caustic curve which is the same as the number 
of the finite limit points.  

In the case of lensing by an elliptical projected mass distribution 
(zeroth approximation of a galaxy lens), the caustic consists 
of one quadroid (or astroid), and the fifth image of a quasar lensed by a 
galaxy is more likely to be missing due to the brightness of the sizeable 
lensing galaxy. However, the color difference of the lensing galaxy and the 
lensed quasar may reveal the small fifth image in high resolution 
images\footnote{PG 1115+080, B1608+656, and B1938+666  
(Fig. 2 of Kochanek et al. 2000) are caustic covering
lensing events and one can infer from their ring images the characteristic 
assemblage of the image domains around critical curves. The image domain
configurations depend on the lensing mass distributions, and the apparent 
(image) rings depend also on the source position with respect to the caustic.
}.
The position of the fifth image in relation to the image ring will be 
one of the clinching constraints in the reconstruction of the image
domains and so the lens characteristics. Draw a circle of radius 
$\sqrt{0.3}= 0.548$ centered at the lens position (marked $\times$) of the
smaller ``thick Einstein ring" (arrowed) in figure \ref{fig-idomain2} and
note that the circle goes through just about the middle of the image 
ring donut. The circle is the Einstein ring of the lensing mass $0.3$ as 
a single lens. Thus, one can more or less read off the lensing mass from 
the image ring. The apprearance of the observed image ring of a host galaxy
of a quasar depends on the position of the source with respect to the caustic,
of course, but the distribution of the bright core images of the quasar
offers information on the point source position.     
We will discuss the image domains, ``thick Einstein rings", 
and Einstein rings of extended mass distribution lenses 
elsewhere \citep{extended}.

\section{ A Conjecture on the Number of Images}

We have meandered through the properties of the caustic curve, the image
curves of the caustic curve, image domains, the critical curve and smoothly 
rotating Jacobian eigenvectors on the critical curve, and the limit points
and lens positions as the ``sinks" and ``sources" of the potential defined
by $|\kappa(z)|$. The number of images jumps across the caustic curve 
because of the pairs of positive and negative images across the critical 
curve. Thus, if we refer to the image curves of the caustic curve other than 
the critical curve as precaustic curves, the critical curve is ``padded"
from both sides by the positive and negative image domains defined by the
precaustic curves. Precaustic curves are tangentially attached to the critical 
curve at designated precusps, and their $J$ values are either positive 
(positive precaustic curve) or negative (negative precaustic curve) except 
at the precusps where they are tangent to the critical curve. The  number 
of positive images of a given caustic is given by the number of positive
image domains that are mapped to the caustic domain by the lens equation,
and figure \ref{fig-idomain} and \ref{fig-idomain2} illustrate that each
positive image domain defined by a positive precaustic curve is associated
with a finite limit point. So, a conjecture follows.          

\begin{enumerate}
\item
{\it Conjecture: \ 
  The number of positive images of an $n$-point lens is bounded by the number
  of finite limit points $2(n-1)$ for $n > 1$.}  When $n=1$, there is one 
  limit point at $\infty$ and a point source has one positive
  image as is well known.
\item  
{\it Corollary: \ 
  The total number of images of an $n$-point lens can not exceed 
  $5(n-1)$ when $n>1$.} When $n=1$, there are two images. 
 \item   
{\it 
  The maximum number of images of an $n$-point lens can not be less
  than $3(n+1): n \ge 3$.}
\end{enumerate}
The corollary follows from the conjecture because there are always 
$(n-1)$ more negative images in an $n$-point lens. In order to prove the 
third item, we consider a lens consisting of $n$ equal masses 
equally spaced on a circle of radius $a$ can produce an image at every 
distinct limit point. We choose a source at the center of circle to 
maximally utilize the symmetry and examine the cases of $n = 2^k$.  
The lens equation (\ref{eqLens}) becomes
as follows where $z_j \equiv z - x_j$. 
\opeqn
 z = \sum_1^n {1\over n} {1\over \bar z_j} \equiv \bar F_n
\label{eqSym}
\cleqn
For $n=2$, we let $x_{1,2} = \pm a$, and  
\opeqn 
 \bar F_2 = {\bar z\over \bar z^2 - a^2} \ . 
\cleqn
For $n=4$, we add two equal masses at $x_{3,4} = \pm i a$, and 
\opeqn
 \bar F_4 = \bar F_2(a^2) + \bar F_2(-a^2) 
     = {\bar z^3 \over \bar z^4 - a^4} \ . 
\cleqn
The pattern is clear: $\bar F_8 = \bar F_4(a^4) + \bar F_4(-a^4)$, etc, 
and
\opeqn
 \bar F_n = =  {\bar z^{n-1}\over \bar z^n - a^n} \ . 
\cleqn
The limit points $z_\ast$ are the zeros of $\partial F_n$.  
\opeqn
 0 = z^{n-2} (z^2+(n-1)a^n)
\cleqn
The center $z_\ast =0$ is an (n-2)-th order zero and satifies the
lens equation (and so an image).  Other $n$ zeros are on the circle 
of radius $r_\ast = (n-1)^{1/n} a: r_\ast < a $.      
\opeqn 
  z_\ast = r_\ast e^{\pi\over n}, \ r_\ast e^{3\pi\over n}, \ ... \ 
           r_\ast e^{(n-1)\pi\over n} 
\cleqn
They are images when $a = a_n$.
\opeqn
  a_n =  n^{-{1\over 2}} (n-1)^{n-2\over n}  
\label{eqAvalue}
\cleqn
As $n $ becomes large, $a_n $ converges to $1$. 
Since the $n$-point ``necklace" lens has $(n+1)$ distinct limit points,
the source at the center produces no less than $(n+1)$ positive images,
or equivalently no less than $3n+1$ images in total. One image is
at the center, and $3n$ images are on three circles of radius
$r_\ast$ ($n$ positive images), $r_< < a$ ($n$ negative images), and
$r_> > a$ ($n$ negative images).  In the case of  $n = 4$, $a=0.658$, 
$ r_\ast = \sqrt{3}/2 = 0.866$, $r_< = 0.5$, and 
$r_> = 1/2 + \sqrt{7}/4 = 1.16$. 
We have analysed the cases where $n$ is power of 2, but the (discrete)
axial symmetry of the ``necklace" lenses gurantees that $3n+1$ is the
lower bound of the maximum total number of images of arbitrary $n$-point 
lenses. 

The formula $3n+1$ produces the correct maximum possible number of images 
$10$ for $n=3$ because it happens to be that the two formulae for the number 
of limit points $2(n-1)$ and $n+1$ coincide when $n=3$. \citet{mao97}  
numerically examined some cases of the family of ``necklace" lenses and observed 
that the maximum number of images was $3n+1$. The formula $3n+1$ predicts too 
many images for $n = 2$ and $n=1$ because the number of the limit points is 
less than $(n+1)$ and 2 less images for $n=4$. Here we have shown that
$3n+1$ is the lower bound of the maximum possible number of images of an
$n$-point lens for $n \ge 3$ by investigating source positions that generate 
positive images at the limit points as we did in rh97.


\def\ref@jnl#1{{\rm#1}}
\def\aj{\ref@jnl{AJ}}
\def\apj{\ref@jnl{ApJ}}
\def\apjl{\ref@jnl{ApJ}}
\def\apjs{\ref@jnl{ApJS}}
\def\aap{\ref@jnl{A\&A}}
\def\aapr{\ref@jnl{A\&A~Rev.}}
\def\aaps{\ref@jnl{A\&AS}}
\def\mnras{\ref@jnl{MNRAS}}
\def\prl{\ref@jnl{Phys.~Rev.~Lett.}}
\def\pasp{\ref@jnl{PASP}}
\def\nat{\ref@jnl{Nature}}
\def\sci{\ref@jnl{Science}}
\def\iauc{\ref@jnl{IAU~Circ.}}
\def\aplett{\ref@jnl{Astrophys.~Lett.}}
\def\annrev{\ref@jnl{Ann.~Rev.~Astron.~and Astroph.}}


\begin{figure}
\plotone{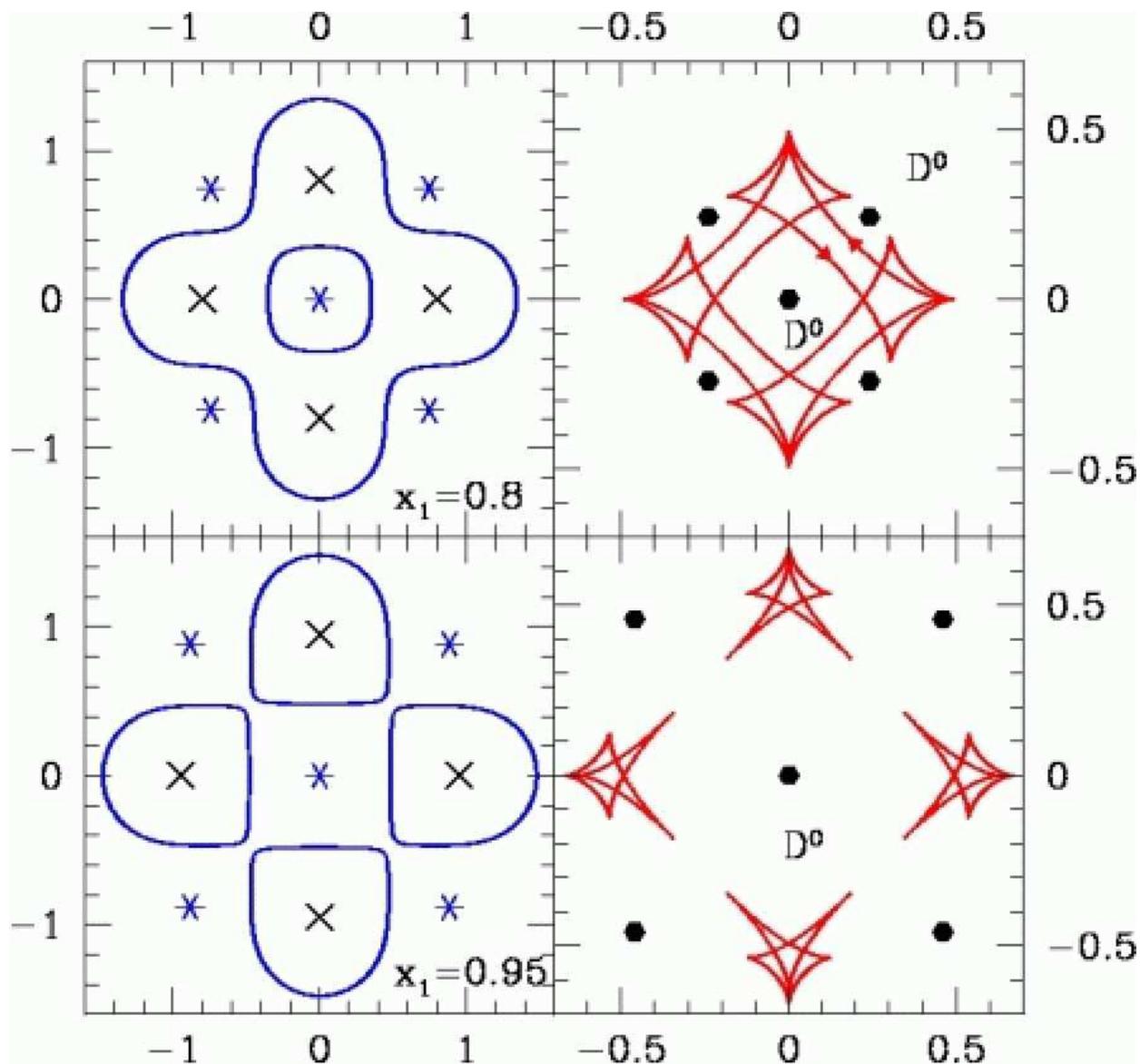}
\caption{ As the lens separation increases, the gravitational interference 
  is fractionized and the ``traffic" of the caustic curves in the central
  region moves in the way the highest degree domain in the center shrinks. 
  ${\cal D}^4$ disappears at $a = 1/\sqrt{2}$, and the central region becomes 
  ${\cal D}^0$ where the number of images is the minimum 5. With increasing
  $a$ ($x_{1,3} = \pm a, \  x_{2,4} = \pm i a$), the caustic curve degenerates 
  into caustic loops directly associated with the lens elements and the loops
  shrink and unwrinkle.   
  \label{fig-sym4-cc1} }
\end{figure}

\begin{figure}
\plotone{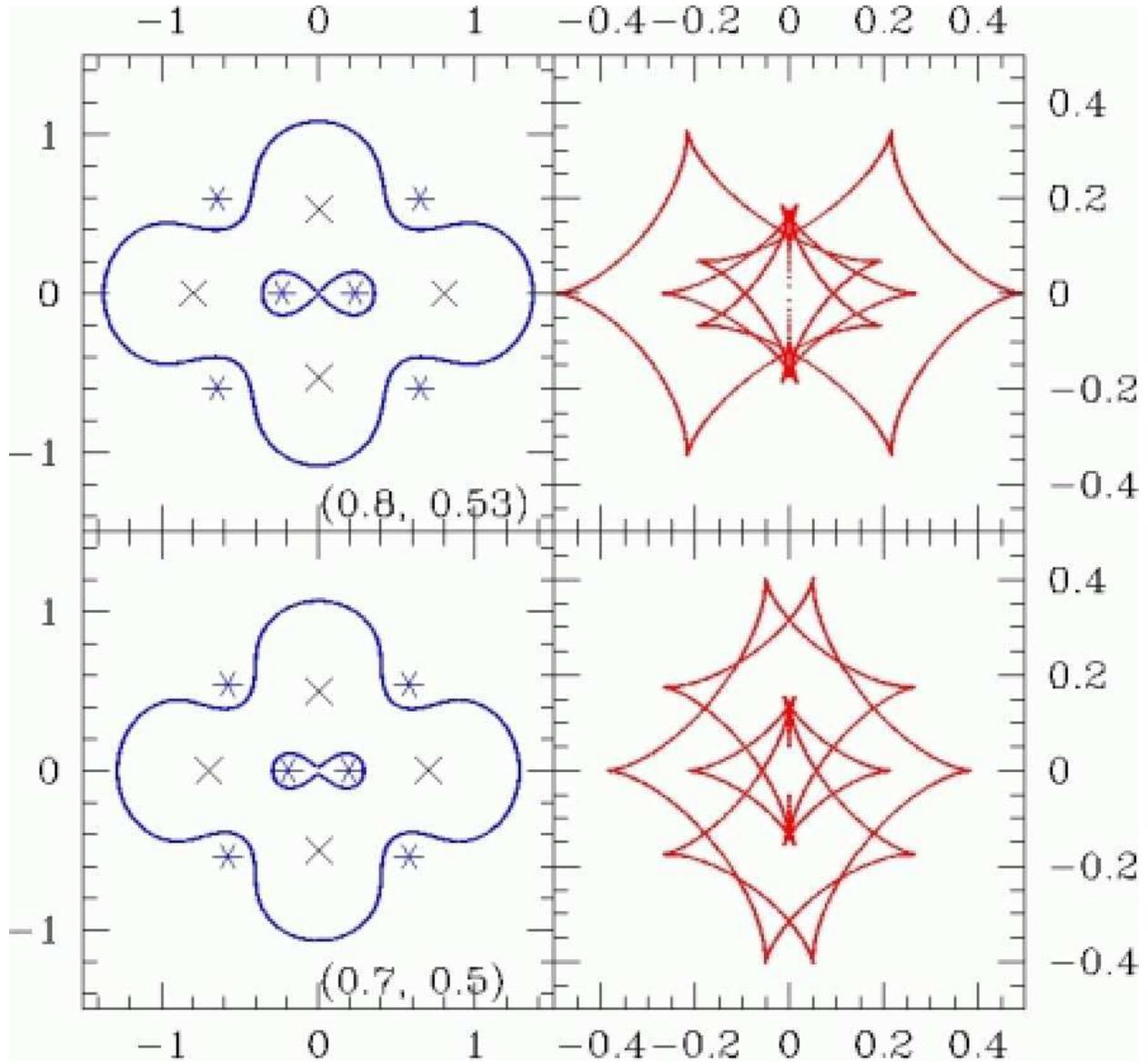}
\caption{ The rhombus lens configuration in figure 2 and figure 3 
 is changed so that the rhombus becomes more regular. The shorter 
 diagonal is extended in the upper panel and the longer diagonal is 
 shortened in the lower panel. The two critical loops around the limit points 
 at the center merge into one, the corresponding caustic loops merge, 
 and the caustic domains ${\cal D}^5$ shrink into the small areas defined 
 by the 8-cusp caustic loops.  The 8-cusp caustics unwrinkle into 4-cusp 
 caustics as the rhombus becomes more regular.  
  \label{fig-sym4-ab1} }
\end{figure}

\begin{figure}
\plotone{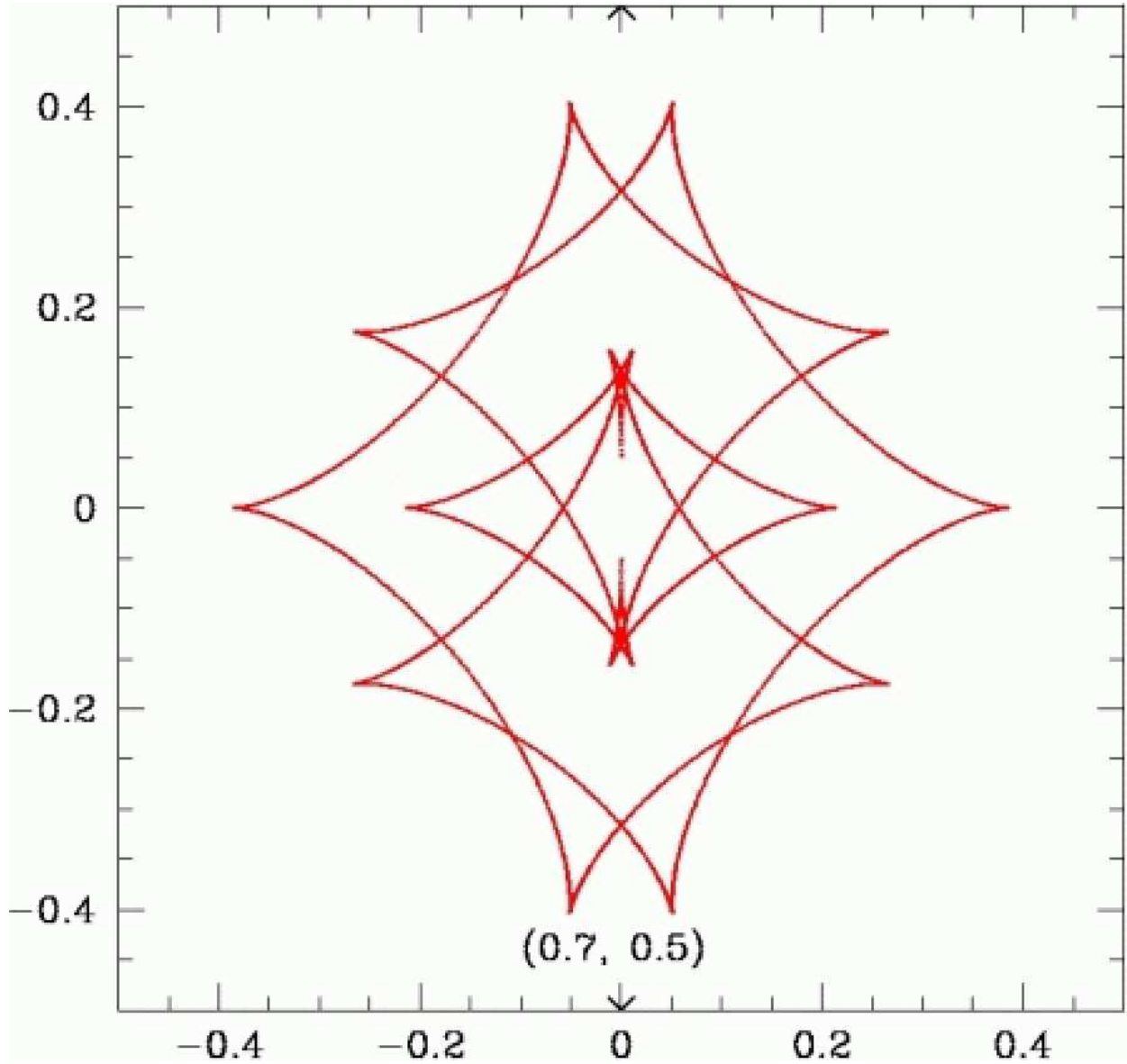}
\caption{ The curves have $e=4$. The critical curve is almost pinching off 
  a critical loop around a limit point ($z_\ast$), and the caustic curve 
  shows a triangular shape segment enclosing the corresponding point 
  $\omega_ast$. The little sharp ``wrinkles" around the opening of the
  triangle are common in $n$-point lenses with $n \ge 3$ because the caustic 
  curve intersects itself. 
  \label{fig-sym3lim} }
\end{figure}

\begin{figure}
\plotone{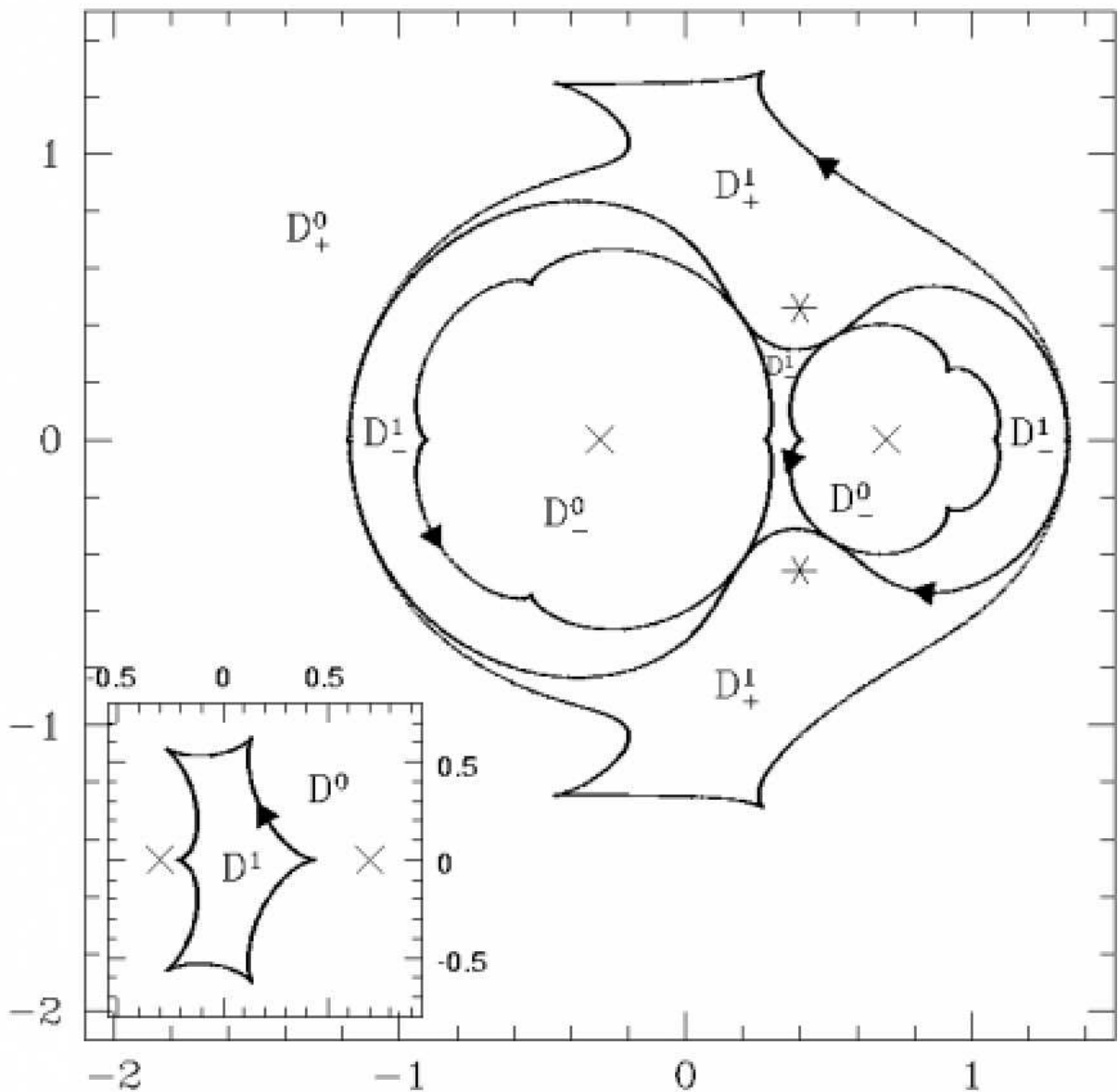}
\caption{ Image domains of a binary lens caustic with six cusps. The arrows
 indicate the orientations of the caustic curve (inset), the image curves of 
 the caustic curve. The positions of the triangle arrows are the source and
 image positions and one can infer the speeds of the image motions along the
 critical curve and precaustic curves. The smooth curve is the critical curve.
 When the source is outside the caustic curve (${\cal D}^0$), there are 3 
 images (2 ${\cal D}^0_-$ and 1 ${\cal D}^0_+$). A source inside the caustic
 curve (${\cal D}^1$) generates 5 images 
(3 ${\cal D}^1_-$ and 2 ${\cal D}^1_+$). Each domain ${\cal D}^1_+$ is 
associated with one finite limit point. The positive domain ${\cal D}^0_+$ 
 is associated with the limit point at $\infty$.
  \label{fig-idomain} }
\end{figure}

\begin{figure}
\plotone{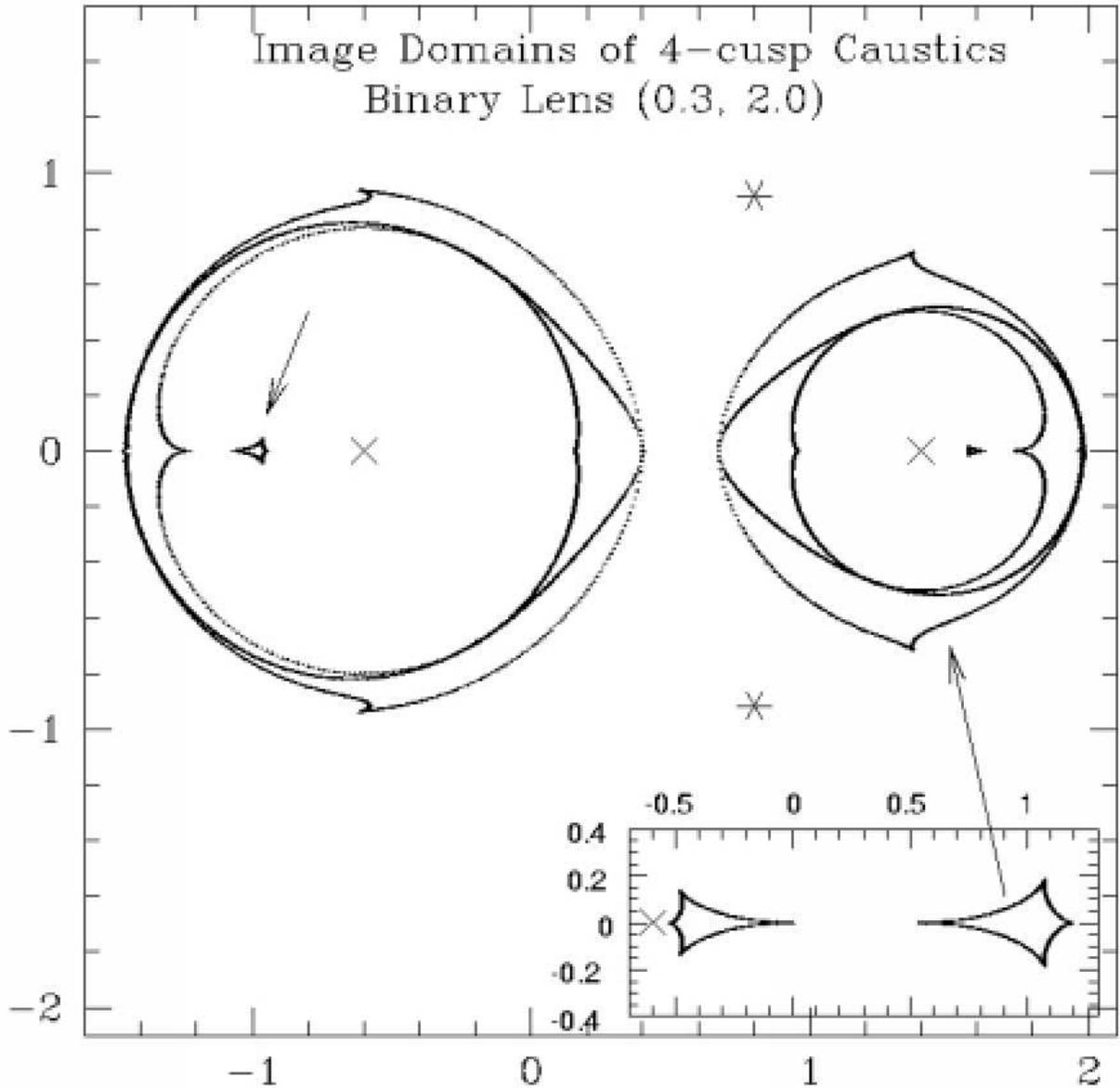}
\caption{ Image domains of a binary lens caustic with four cusps.  
  The interior of each quadroid commands 5 image domains. The ``thick
  Einstein rings" (image rings) are made of 4 image domains (2 positive
  and 2 negative) as indicated by the long arrow. The fifth image in the 
  small (negative) image domain indicated by the short arrow has a high
  probability to be missing in an observation. The smooth curve is the  
  critical curve. Despite the oval shapes of the critical loops, the 
  ``thick Einstein rings" are more or less circular annuluses. The Einstein
  ring of each lens element as a single lens traces more or less the middle
  of the corresponding image ring as centered at the lens position ($\times$).  
  \label{fig-idomain2} }
\end{figure}

\end{document}